\documentclass[conference, 10pt]{IEEEtran}
\usepackage{amsthm}

\newtheorem{Exam}{Example}

\usepackage{enumitem}
\newcounter{problem}
\newcounter{subproblem}[problem]
\newenvironment{problem}{\refstepcounter{problem}{\bfseries Problem~\theproblem}}{}

\usepackage{hyperref}
\usepackage{cleveref}

\usepackage{mathtools}

\usepackage{color}
\usepackage{xcolor}
\usepackage{colortbl}
\definecolor{purple}{RGB}{139, 0, 139}
\usepackage{manfnt}
\usepackage{cleveref}
\usepackage{url}

\usepackage{caption}
\usepackage{subcaption}

\newif\iftodo   
\todotrue
\newif\iftodoshort  
\todoshortfalse
\ifCLASSINFOpdf
  \usepackage[pdftex]{graphicx}
   \DeclareGraphicsExtensions{.pdf,.jpeg,.png,.tiff}
\else
   \usepackage[dvips]{graphicx}
   \DeclareGraphicsExtensions{.eps,.ps}
\fi
\usepackage{amssymb,amsmath}
\usepackage[mathscr]{euscript}
\usepackage{bbm}
\usepackage{dsfont}

\DeclareMathOperator*\argmax{arg \, max}		
\DeclareMathOperator*\maximize{max.}		
\usepackage{algorithmic}
\usepackage[ruled,vlined,commentsnumbered]{algorithm2e}

\newcommand{\Rmnum}[1]{\uppercase\expandafter{\romannumeral #1}}
\newcommand{\rmnum}[1]{\lowercase\expandafter{\romannumeral #1}}

\newcommand{\field}[1]{\mathbb{#1}}

\newcommand{\emenge}[1]{\mathscr{#1}}
\newcommand{\set}[1]{\mathscr{#1}}

\newcommand{\RN}{{\field{R}}_{+}}
\newcommand{\RP}{{\field{R}}_{++}}

\newcommand{\Ts}{{\emenge{T}}}
\newcommand{\Ss}{{\emenge{S}}}

\newcommand{\Ps}{{\emenge{P}}}

\newcommand{\Ds}{{\emenge{D}}}

\newcommand{\ve}[1]{\boldsymbol{\mathbf{#1}}}

\newcommand{\vx}{\ve{x}}

\newcommand{\vp}{\ve{p}}

\usepackage[utf8]{inputenc}

\newcommand{\ul}{\text{UL}}
\newcommand{\dl}{\text{DL}}

\usepackage{hyperref}
\usepackage{amsthm}
\usepackage{amssymb,amsmath}
\usepackage[mathscr]{euscript}

\newcommand{\cosl}[1]{}
\newcommand{\resl}[1]{}

\usepackage[draft,footnote,nomargin]{fixme}
\newcommand{\fnql}[1]{}
\newcommand{\fnsv}[1]{}

\begin{document}
%
\title{Resource Scheduling for Mixed Traffic Types with Scalable TTI in Dynamic TDD Systems}
\author{
\IEEEauthorblockN{Qi Liao\IEEEauthorrefmark{1}, Paolo Baracca\IEEEauthorrefmark{1}, David Lopez-Perez\IEEEauthorrefmark{2} and Lorenzo Galati Giordano\IEEEauthorrefmark{2}}
\IEEEauthorblockA{\IEEEauthorrefmark{1}Nokia Bell Labs, Stuttgart, Germany\\ 
\IEEEauthorrefmark{2}Nokia Bell Labs, Dublin, Ireland\\
Email: \{qi.liao, paolo.baracca, david.lopez-perez, lorenzo.galati\_giordano\}@nokia-bell-labs.com}
}

\maketitle
\begin{abstract}

This paper analyses the performance benefits of a user-centric scheduling approach, 
exploiting the flexibility of both dynamic time division duplex (TDD) and a variable transmission time interval (TTI), 
where the downlink to uplink  ratio and TTI duration can be adapted to the traffic load. 
The formulation of the joint optimisation problem takes into consideration the individual requirements of each single user in terms of sustainable latency and desired throughput, 
thus implementing a real user-centric scheduling approach. 
Moreover, the developed solution is evaluated in a scenario with mixed traffic types, 
mobile broadband (MBB) and mission critical communications (MCC), 
showing remarkable performance enhancement of the proposed scheme  over baseline dynamic TDD schemes with a fixed TTI 
in terms of both  achievable throughput of the MBB users and guaranteed latency for the MCC users.\\    

\end{abstract}

\section{Introduction}

One of the main drivers of next generation mobile networks is the future heterogeneity of services and applications~\cite{ngmn_whitepaper}.
Besides the support for mobile broadband (MBB), 
the fifth generation (5G) systems should also manage machine type of communications (MTC), 
which are mostly characterised by small packet transmissions, 
and have very different requirements than MBB traffic.
For example, two representative use cases of MTC are massive machine communications (MMC) and mission critical communications (MCC)~\cite{pedersen_cm16}.
While MMC applies to scenarios with a huge number of low-cost devices transmitting only sporadically with quite relaxed latency requirements but strict energy constraints, 
MCC refers to scenarios where nodes have low to moderate throughput demands but stringent latency and reliability demands~\cite{durisi_arxiv16}.
This traffic diversity calls for a thorough reappraisal of contemporary wireless network technologies, 
and the change from a base station (BS)-centric to a user equipment (UE)-centric networking paradigm,
where only the necessary resources, no more, no less, are allocated to each UE to satisfy its throughput, latency and energy requirements.


Recent studies have shown that time division duplex (TDD) represents a more flexible option than frequency division duplex (FDD) system to manage this traffic heterogeneity and realise such UE-centric networking~\cite{lahetkangas_icc14}.
In this line, the third generation partnership project (3GPP) has already introduced a number of semi-static TDD configurations in the last LTE releases
that can be dynamically selected by the BS
to deal with traffic burstiness~\cite{3gpp_tr36828}.
Moreover, the 3GPP has also commenced to consider the support for shorter and variable  time transmission intervals (TTIs),
which will 
reduce latency at the physical and medium access control layers, and
further help to provide a tailored amount of resources to UEs,
while avoiding any waste~\cite{pedersen_cm16}.

Dynamic TDD allows a BS to dynamically adapt the downlink (DL) to uplink (UL) ratio to the current traffic load. 
It has been shown in several works that BS clustering, a method that groups the cells to adopt the same DL to UL ratio when they are characterized by high interference \cite{ding_icc14} and have similar traffic profile and buffer size \cite{baracca_vtc16}, can cope with DL-to-UL and UL-to-DL interference and provide good throughput for MBB users at a reasonable complexity.
%
However, these works do not take into account different services classes. 
Motivated by the flexible frame structure in 5G, \cite{pedersen2015flexible} proposed to fulfil the strict delay constraints for MCC users by dynamically selecting the TTI length in FDD system, using reverse calculation based on the delay budgets. Along similar lines, in \cite{levanen2014radio} the authors apply the variable frame duration to achieve the ultra-low latency in millimeter-wave communication. Both works use the strategy that short TTI is selected first when MCC users exist in the system, followed by long TTI lengths when reaching steady state operation. However, such schemes always provide priority to the MCC users to fulfil their latency constraints, while sacrificing the throughput performance of the others.

In this paper, we aim at developing a true user-centric approach that provides a flexible tradeoff between mixed types of services to meet their specific requirements in both UL and DL for dynamic TDD systems. 
%
To the best of the authors' knowledge, this is the first work that develops a dynamic TDD framework for a scenario with mixed types of services (where UEs generate either MBB or MCC traffic in both UL and DL), 
considering scalable TTI lengths and individual UE requirements in terms of both throughput and latency.
Our proposed user-centric approach decides for each scheduling time: 
\emph{a)} the duplexing mode, i.e., either DL or UL, 
\emph{b)} the TTI length, and 
\emph{c)} the UEs to be served and the resources allocated in each TTI. 
Numerical results show that the proposed scheme significantly outperforms dynamic TDD schemes with a fixed TTI in terms of both throughput provided to the MBB UEs and latency guaranteed to the MCC UEs. When compared to the schemes that always admit high priority to MCC UEs, our proposed scheme can achieve comparably low latency for MCC UEs, while providing good throughput also for the MBB UEs.

The rest of the paper is organised as follows. In Section~\ref{system_model}, the system model is described together with the correspondent notation. In Section~\ref{problem_solution}, the problem formulation is introduced with an explanation of the different utility functions and the correspondent optimal solution. Finally, in Section~\ref{num_results}, the numerical results are presented, followed by the concluding remarks drawn in Section~\ref{conclusions}.    


\section{System Model} \label{system_model}

We will use the following notation throughout the paper.
Bold capital letters and bold lowercase letters denote matrices and vectors, respectively,
while $\set{X}$ denotes a set, 
and $\vx\succeq 0$ implies that $x_i\geq 0$ for all components $i$. 
Moreover, the set of non-negative real numbers and the set of positive real numbers are denoted by $\RN$ and $\RP$, respectively. 
The cardinality of set $\set{X}$ is denoted by $|\set{X}|$, 
and the Cartesian product of two finite sets $\set{X}$ and $\set{Y}$ is denoted by $\set{X}\times\set{Y}$.  

\begin{figure}[t]
  \centering
 \includegraphics[width=.45\textwidth]{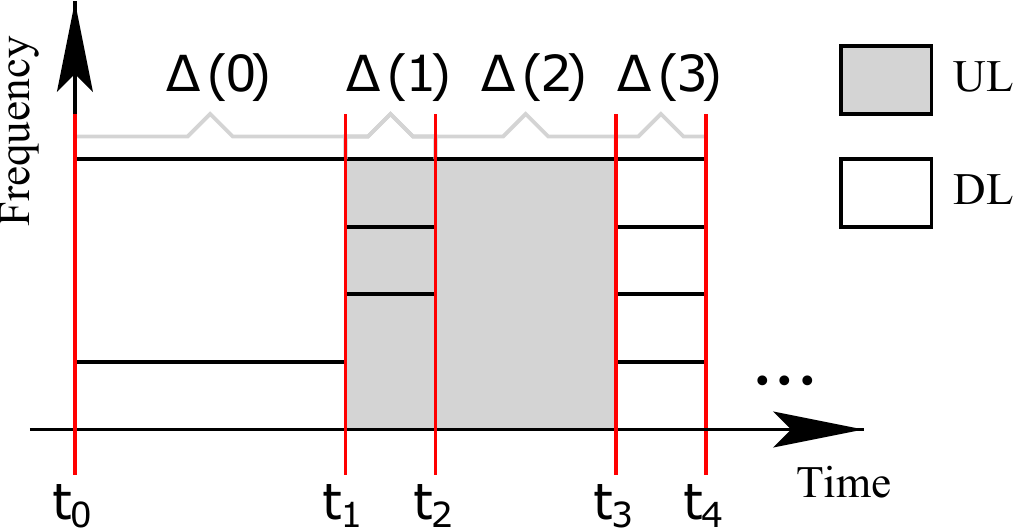}
  \caption{Dynamic TDD frame structure with scalable TTI.}
  \label{fig:scalableTTI}
	\vspace{-1em}
\end{figure}

We consider a single cell scenario 
where we assume that the system operates in a dynamic TDD mode,
and that the average received co-channel inter-cell interference can be estimated. 
Under this setup, 
let the TTI be discrete, 
and indexed by $n\in\{0,1, \ldots\}$, 
where the length of the TTI $\Delta(n)$ is scalable as shown in Fig.~\ref{fig:scalableTTI} and can be chosen from a finite set of $M$ lengths $\Ts = \{\Delta_i: i=1, \ldots, M\}$. 
In addition, let $t_n$ denote the time point at which the $n$th TTI begins  
(which is also the time point at which the $(n-1)$th TTI finishes if $n>0$). 


At the $n$th TTI, 
the set of existing services (including both uplink and downlink)\footnote{
Here a service refers to the communication between two nodes that occurs during the span of a single connection. 
For example, a service  of web browsing (in downlink) starts when a user requests to connect to one URL and ends when all information from that URL is displayed.} 
is denoted by $\Ss(n):=\{s: s = 1,  \ldots, S(n)\}$. 
Now and hereafter, 
we omit the index $n$ when we can do so without any ambiguity. 
The set $\Ss$ comprises the subsets $\Ss^{(d)}$, $d\in\Ds$, 
where $\Ds :=\{\ul, \dl\}$ denotes the duplexing mode of uplink or downlink. 
Since a transmission link can be either uplink or downlink in the dynamic TDD system, 
we separate the uplink and downlink services such that $\Ss^{(\ul)}\cup\Ss^{(\dl)} = \Ss$ and $\Ss^{(\ul)}\cap\Ss^{(\dl)} = \emptyset$,
where $\left|\Ss^{(d)}\right| = S^{(d)}$ for $d\in\Ds$ denotes the cardinality of set $\Ss^{(d)}$.

Our target is to construct a scheduler 
which selects at each TTI, 
a duplexing mode $d\in\Ds$ and a TTI length $\Delta\in\Ts$,  
and schedules {\it one or more services} in the selected duplexing mode \footnote{Since we conduct our study in a single cell scenario in this paper, TTI length and duplexing mode are selected regardless the configuration of the neighboring cells. Frame alignment and inter-cell inter-mode interference (between uplink and downlink) will be considered in the follow-up work.}. 
Under this setup, 
let $\vp:=(p_1, \ldots, p_S)^T\in\Ps:=[0,1]^{S}$ denote the fraction of frequency resources assigned to services,
which should satisfy the following resource constraint $\sum_{s\in\Ss(n)} p_s(n)\leq 1$.  

\subsection{Achievable Rate}
\label{eqn:rate}

We assume that the channel state remains the same within a TTI (regardless of the TTI lengths), 
and that the coherence time of the channel lasts in general more than one TTI. 
Further, we assume that signal to interference plus noise ratio $\gamma_s(n)$ can be estimated for each transmission link of service $s$  and for each TTI $n$
(perfect knowledge of  transmission power, channel states, received interference and noise power).

In the following, we introduce the achievable rate considering the impact of the control signalling overhead.
In fact, the TTI length is configured as a multiple of a number of orthogonal frequency division multiplexing (OFDM) symbols,
and the signalling overhead is fixed.
Thus, although some services can be scheduled with a short TTI to fulfil their strict latency requirement, 
this short TTI comes at the expense of higher control signalling cost and lower capacity for the services requiring higher data rates~\cite{pedersen_cm16}.

%

Taking the signalling overhead into account, 
the achievable data rate of service $s$ at the $n$th TTI is given by
\begin{align}
	& r_s(n)  := r_s\left(p_s(n), d(n), \Delta(n)\right) \nonumber\\
 	& =  \mathds{1}_{\{s\in\set{S}^{(d(n))}(n)\}}  \psi(\Delta(n)) p_s(n)  B \log\big(1+\gamma_s(n)\big)
	\label{eqn:datarate_excludecontrol}
\end{align}
where 
$\mathds{1}_{\left\{\text{A}\right\}}$ is the indicator function, 
which equals to $1$ if the event $\text{A}$ occurs, 
and zero otherwise,  
$\set{S}^{(d(n))}(n)$ is the set of services at duplexing mode $d(n)$ at the $n$th TTI 
and $B$ denotes the total bandwidth.
Moreover, $\psi(\cdot)$ is a function indicating the ratio of the effective time to transmit the payload. 
For example, if we assume that a fixed time duration $\tau$ is required for the control signal transmission on the dedicated channel for any TTI length,
where $\tau$ is smaller than the shortest TTI length,
we can then define $\psi: \Ts\to(0,1]: x\mapsto (x-\tau)/x$.


\subsection{Time-Varying Service Demand}
\label{subsec:demand}

Let $t_s'$ denote the arriving time of service $s$, 
and $n_s'$ denote the index of the TTI that contains the time instant $t_s'$, 
i.e., $t_s'\in [t_{n_s'}, t_{n_s'+1})$.
Each new incoming service $s$ arrives with a traffic demand of $\nu_s(n_s')$ (in bits) and a latency constraint of $l_s(n_s')$ (in seconds). 
As the scheduler allocates resources to serve it, 
the remaining service demand decreases if the data transmission is successful. 
Note that service $s$ cannot be served before the next TTI (with the index $n_s'+1$) begins. 

By the end of the $n$th TTI with $n>n_s'$, 
the remaining traffic demand of service $s$ is given by 
\begin{equation}
	\nu_s(n) = \nu_s(n-1) - \Delta(n)r_s(n).
	\label{eqn:remainTrafficDemand}
\end{equation}

It is desired that the allocated resources just meet the traffic demand, 
but do not exceed it, otherwise, it is a waste of resources. 
Thus, we define an additional constraint to avoid negative values of $\nu_s(n)$, 
expressed as
\begin{equation}
	\nu_s(n-1) - \Delta(n)r_s(n)\geq 0.
	\label{eqn:NonnegativeQueue}
\end{equation}

The remaining latency constraint for $n >n_s'+1$,
i.e., the maximum remaining time to fulfil the traffic demand, 
is written as   
\begin{equation}
	l_s(n) = 
	\left(l_s(n-1) - \Delta(n)\right)^{+}.
	\label{eqn:remainLatency}
\end{equation}
where $(\cdot)^{+}:=\max\{\cdot, 0\}$. 
For $n = n_s'+1$, 
we have that $l_s(n) = (l_s(n-1) - (t_{n_s'+1}-t_s'))^{+}$, 
since service $s$ cannot be served before the $(n_s'+1)$th  TTI begins.
Note also that $l_s(n)=0$ implies that the latency constraint has expired. 

Finally, a service is removed from the system, 
if the remaining traffic demand yields $\nu_s(n)=0$.

\section{Problem Formulation and Optimal Solution} \label{problem_solution}

By optimising the variables $\{\vp(n), d(n), \Delta(n)\}$ at the $n$th TTI,
our objective is to design a computationally efficient scheduling algorithm for mixed traffic types,
which can work on a very short time scale and achieve a good tradeoff between heterogeneous performance metrics with respect to traffic demands and latency requirements. 

\subsection{Latency-Aware Cost Function}\label{subsec:PM}

To penalise the undesirable long latency period for the MCC services characterised by strict low latency requirements, 
we introduce a cost function defined as follows.
\begin{align}
	J_s(n)& :=J_s(p_s(n), d(n), \Delta(n)) \nonumber\\
	& =\dfrac{\dfrac{\nu_s(n)}{\overline{R}_s(n)}+\left(\Delta(n)-l_s(n-1)\right)^{+}}{\max\{l_s(n),c_{\min}\}}
	\label{eqn:LatencyMetric}
\end{align}
where $\nu_s(n)$ depending on $(p_s(n), d(n), \Delta(n))$ is given by~\eqref{eqn:remainTrafficDemand}, 
$0<c_{\min}\ll \Delta_{\min}$ is arbitrarily small to prevent divide-by-zero errors, 
and $\Delta_{\min}:=\min_\Delta \set{T}$ is the minimum TTI length. 
At the numerator,
the term $\nu_s(n)/\overline{R}_s(n)$\footnote{
The term $\nu_s(n)/\overline{R}_s(n)$ in \eqref{eqn:LatencyMetric} is designed as the best-case estimate of the time to serve the remaining traffic. 
The myopic optimisation scheme cannot estimate the remaining delay because it depends on the scheduling decision in the future. 
However, we can offer a best-case estimate for the remaining serving time under the optimal condition that a service is scheduled with full resource allocation. 
This relaxes the penalty for the MBB services because there is usually some room to fulfill the demand in the upcoming TTIs if $l_s(n-1)$ is large enough, 
while it raises the cost of MCC services if a long TTI is used, leading to a small value of $l_s(n) = (l_s(n-1)-\Delta(n))^{+}$.}
serves as the {\bf best-case estimate of the time to serve the remaining traffic} by the end of the $n$th TTI, 
with $\overline{R}_s(n)$ defined as the best-case serving rate after the $n$th TTI,
while the term $\left(\Delta(n)-l_s(n-1)\right)^{+}$ is an {\bf offset} to guarantee that assigning a TTI longer than the required latency constraint results in a high cost, 
even if the remaining traffic can be served within the chosen TTI 
(i.e., $\nu(n) = 0$).
At the denominator,
the term $\max\{l_s(n),c_{\min}\}$ is the remaining latency constraint indicating the {\bf maximum required time to serve the remaining traffic} by the end of the $n$th TTI. 
It is important to note that $\overline{R}_s(n)$ can be estimated by 
averaging the maximum achievable rate $B\log(1+\gamma_s(j))$ over the recent time period of successive TTIs $j\in [n_s',n]$. 
To better describe the physical meaning of the cost function in \eqref{eqn:LatencyMetric}, 
we provide a simple numerical example here below.

\begin{Exam}
Assume that at the initial time point, 
a MBB service ($s=1$) and a MCC service ($s=2$) arrive in uplink with the traffic demands $\nu_1(0) = 10^3$ and $\nu_2(0) = 2$ (in bits),  
and that the latency constraints are $l_1(0) = 5$ and $l_s(0)  = 0.25\cdot{10^{-3}}$ (in seconds), respectively. 
Define $c_{\min} = 10^{-6}$. 
Assume that the achievable rate remains the same for the next $10^{-3}$ seconds 
and we have $r_1(1) = 10^4$ and $r_2(1) = 8\cdot 10^3$ (in bit/s) 
if the full bandwidth is allocated (i.e., $p_1=1, p_2 = 0$ or $p_1 = 0, p_2=1$), respectively. 
We consider the following three cases of the configuration for the $1$st TTI. 
\\
1) Configuration $\Delta(1) = 10^{-3}$, $d(1) = \ul$, $p_1(1) = 1$, $p_2(1) = 0$. 
We have  $J_1(1) = \left(\left(10^3-10\right)/10^4+0\right)/\left(5-10^{-3}\right)\approx 0.0198$, 
while $J_2(1)=\left(2/\left(8\cdot{10^3}\right)+0.75\cdot 10^{-3}\right)/10^{-6}=10^3$. 
In this case, using a long TTI and allocating all resources to the MBB service cause high latency cost of service $2$.  
\\
2) Configuration $\Delta(1) = 10^{-3}$, $d(1) = \ul$, $p_1(1) = 0.75$, $p_2(1) = 0.25$. 
We have $J_1(1) = \left(\left(10^3-7.5\right)/10^4+0\right)/\left(5-10^{-3}\right) \approx 0.0199$, 
while  $J_2(1)=\left(0+0.75\cdot 10^{-3}\right)/10^{-6}=750$. 
In this case,  although for service $2$ the remaining traffic is served within one TTI, 
its latency cost is still high due to the violation of the latency constraint. 
\\
3) Configuration $\Delta(1) = {0.25}\cdot{10^{-3}}$, $d(1) = \ul$, $p_1(1) = 0$, $p_2(1) = 1$. 
We have $J_1(1) = \left(10^{3}/10^{4}+0\right)/\left(5- 0.25\cdot{10^{-3}}\right)\approx 0.02$ 
while $J_2(1) =0$. 
Compared to the first configuration, 
$J_1(1)$ increases only slightly due to the loose latency constraint of the MBB service, 
while $J_2(1)$ reduces to zero because the strict latency requirement of the MCC service is fulfilled. 
Thus, the third configuration leads to a lower joint latency cost $J_1(1)+J_2(1)$.
\label{exam:costFunct}
\end{Exam}

\subsection{Utility-Based Resource Allocation Problem}
\label{subsec:probem_instan}

In this section, 
we formulate a utility-based optimisation problem,  
where the objective metric with the throughput utility added and the latency-aware cost function subtracted is maximised. 
Omitting the TTI index $n$ for convenience, 
the objective function to maximise is defined as
\begin{align}
 	& U(\vp, d, \Delta) \nonumber\\
	:= & \sum_{s \in \Ss} \Big(\alpha_s u_s\left(r_s(d, \Delta, p_s)\right) - \beta_s J_s(d, \Delta, p_s)\Big) 
	\label{eqn:utilityDef}
\end{align}
where the concave utility function $u_s  :  \RN\to\RN \nonumber:x\mapsto \log\left(1+x\right)$ is applied to achieve throughput fairness among the services, 
and $\alpha_s\in\RN$ and $\beta_s\in\RN$ for $s\in\Ss$ are the service-specific weight factors to give different priorities to services with different rate or latency requirements, respectively. 
Such  service-specific weight factors are fed back from the UE to the BS, 
and they are the mean to achieve a user-centric networking.
For example, 
a higher $\alpha_s$ can be defined for the MBB services with high throughput demands, 
or a higher $\beta_s$ can be defined for the MCC services with strict latency requirements. 
The optimisation of the values of $\alpha_s$ and $\beta_s$ for each service is outside the scope of this paper.



The optimisation problem for each TTI over a set of variables $(\vp, d,\Delta)$ is written as

\vspace{1em}

\begin{problem}[Joint optimisation problem]
\begin{subequations}
\label{eqn:problem_1}
\begin{align}
\maximize\limits_{\vp, d, \Delta} \quad & U(\vp, d, \Delta)\\
\mbox{s.t. } 
 & \eqref{eqn:datarate_excludecontrol}, \eqref{eqn:remainTrafficDemand}, \eqref{eqn:NonnegativeQueue}, \eqref{eqn:remainLatency}, \eqref{eqn:LatencyMetric} \mbox{ and }  \eqref{eqn:utilityDef}\\
&  d\in\Ds, \Delta\in\Ts\\
& \vp\succeq 0, \sum\limits_{s\in\Ss^{(d)}} p_s \leq 1, \ \sum\limits_{s\in\Ss\setminus\Ss^{(d)}} p_s = 0  \label{eqn:ULorDL_1}
\end{align}
\end{subequations}
where \eqref{eqn:ULorDL_1} guarantees that uplink services and downlink services cannot be served at the same TTI in the same cell.   
\label{prob:JointoptimisationProblem}
\end{problem}

Knowing that $|\Ds| = 2$ and assuming that we only have limited choice of TTI lengths with $|\Ts| = M$, 
in order to simplify the joint optimisation problem over the variables $(\vp, d,\Delta)$, 
we can optimise $\vp$ in Problem~\ref{prob:JointoptimisationProblem} for each combination of fixed parameters $(d', \Delta')$, 
and find the optimum solution to the following subproblem.
 
\vspace{1em}

\begin{problem}[Subproblem with fixed $(d', \Delta')$]
\label{eqn:problem_2}
\begin{align}
\vp'(\Delta', d') & = \argmax_{\vp} U(\vp, d', \Delta') \label{eqn:OptUtility_prob_2}\\
\mbox{s.t. } 
& \eqref{eqn:datarate_excludecontrol}, \eqref{eqn:remainTrafficDemand}, \eqref{eqn:NonnegativeQueue}, \eqref{eqn:remainLatency}, \eqref{eqn:LatencyMetric},  \eqref{eqn:utilityDef} \mbox{ and } \eqref{eqn:ULorDL_1} \nonumber
\end{align}
where $(d, \Delta)$ in the constraints are replaced by $(d', \Delta')$. 
\end{problem}

Then, the optimum solution to the general Problem~\ref{prob:JointoptimisationProblem}, 
denoted by $(d^{\star}, \Delta^{\star}, \vp^{\star})$, 
can be obtained by searching for the maximum utility $U$ over the solutions to Problem \ref{eqn:problem_2} with respect to every combination of $(d, \Delta)$, 
i.e.,
\begin{align}
	(d^{\star}, \Delta^{\star}) &= \argmax_{d'\in\Ds, \Delta'\in\Ts} U\left(\vp'(d', \Delta'), d', \Delta'\right)\label{eqn:optimum_d_and_Delta}\\
	p^{\star} & = p'(d^{\star}, \Delta^{\star}) \label{eqn:optimumP}
\end{align}
where $U\left(\vp'(d', \Delta'), d', \Delta'\right)$ in \eqref{eqn:optimum_d_and_Delta} is derived from the optimum solution to Problem~\ref{eqn:problem_2}, 
and $p'(d^{\star}, \Delta^{\star})$ is the optimum solution of $\vp$ with respect to the optimum parameters $(d^{\star}, \Delta^{\star})$ found by~\eqref{eqn:optimum_d_and_Delta}.

\subsection{Solution to the Optimisation Problem}
\label{subsec:probem_solution}

Due to the limited number of the combinations of $\Ds$ and $\Ts$ with $|\Ds\times \Ts| = 2M$, 
we can find \eqref{eqn:optimum_d_and_Delta} and \eqref{eqn:optimumP} by exhaustive search with the computational complexity $\mathcal{O}(2M)$.  
Then, the remaining task is to solve Problem~\ref{eqn:problem_2}.

With fixed $d'$, 
\eqref{eqn:ULorDL_1} implies that $p_s = 0$ for all $s\in\Ss\setminus\Ss^{(d')}$. 
To optimise $p_s$ for $s\in\Ss^{(d')}$, 
let us write the equivalent problem of Problem~\ref{eqn:problem_2} here below. 

For convenience's sake, 
we denote the vector of fraction of resources assigned to the services with duplexing mode $d'$ by $\vx:=(p_s: s\in\Ss^{(d')})^T\in[0,1]^{S^{(d')}}$. 
The achievable rate in \eqref{eqn:datarate_excludecontrol} and the cost function in \eqref{eqn:LatencyMetric} are simply linear and affine functions of $x_s$, respectively.
\begin{subequations}
	\label{eqn:simple_R_J_Funct}
	\begin{align}
	r_s(x_s) & := r_s(x_s(n)) = A_s x_s\\
	J_s(x_s) & := J_s(x_s(n)) = B_s + C_s x_s\\
	\mbox{where } &  \nonumber\\
	A_s & := A_s(n) = \psi(\Delta')B\log(1+\gamma_s(n))\\
	B_s & := B_s(n) = \dfrac{\dfrac{\nu_s(n-1)}{\overline{R}_s(n)} + (\Delta'-l_s(n-1))^{+}}{\max\{l_s(n),c_{\min}\}}\\
	C_s & := C_s(n) = -\dfrac{\Delta'A_s}{\overline{R}_s(n)\max\{l_s(n),c_{\min}\}} \label{eqn:Cs}
	\end{align}
\end{subequations}

Moreover, we define the constraint set 
\begin{subequations}
	\label{eqn:setX}
	\begin{align}
	\set{X} & := \{\vx\succeq 0: x_s\leq D_s,  s\in \Ss^{(d')}\} \label{eqn:constraintSetX_1}\\
	\mbox{with } D_s &:=D_s(n)=\frac{\nu_s(n-1)}{\Delta'A_s(n)} \label{eqn:constraintSetX_2}
	\end{align}
\end{subequations}
where~\eqref{eqn:constraintSetX_2} is derived from constraint~\eqref{eqn:NonnegativeQueue}.

In addition, by taking only $s\in\Ss^{(d')}$ into account, 
the objective function in~\eqref{eqn:utilityDef} can be simplified as
\begin{equation}
	V(\vx) = \sum_{s\in\Ss^{(d')}} \Big( \alpha_s u_s\left(r_s(x_s)\right) - \beta_s J_s(x_s) \Big). 
	\label{eqn:modifiedUtility}
\end{equation}

Using~\eqref{eqn:simple_R_J_Funct}, \eqref{eqn:setX} and \eqref{eqn:modifiedUtility},  
the optimisation problem equivalent to Problem \ref{eqn:problem_2} is written in below.

\begin{problem}[Primal problem]
	\label{prob:problem_3}
	\begin{equation}
		V^{\star}:=\max_{\vx\in\set{X}} V(\vx), \mbox{ s.t. } \sum_{s=1}^{S^{(d')}} x_s\leq X_{\max}. 
		\label{eqn:problem_3} 
	\end{equation}
	where the constraint set $\set{X}$ is given in \eqref{eqn:setX}, 
	and $X_{\max}:=\min\{1, \sum_s D_s\}$ is obtained from the individual resource constraint $x_s\leq D_s$ 
	(such that the allocated resource is just sufficient to finish the remaining traffic demand and no resource is wasted) 
	and the sum resource constraint $\sum_s x_s\leq 1$.
\end{problem}

It is easy to verify that $V(\vx)$ is concave in $\vx$, 
because 
\emph{i)}  $u_s\circ r_s$, 
as the composition of a concave and monotone non-decreasing function and a concave function, 
is concave~\cite[pag. 84]{boyd2004convex};
\emph{ii)}  $J_s$, 
as affine function, 
is both convex and concave and the sum of concave functions is concave~\cite[pag. 79]{boyd2004convex}. 
Note that the constraint set defined by $\set{X}$ and $\sum_s x_s\leq X_{\max}$ is also convex. 
Thus, Problem~\ref{prob:problem_3} is a convex optimisation problem, 
which can be solved by looking at the dual formulation 
where there is no duality gap.

Define the {\it Lagrangian} $L(\vx, \lambda)$ for the primal problem in ~\eqref{eqn:problem_3} and the {\it dual function} respectively by 
\begin{align}
	L(\vx, \lambda) &  = V(\vx)+\lambda(X_{\max}-\sum_s x_s) \label{eqn:Lagrangian}\\
	L(\lambda) & := \max_{\vx\in\set{X}} L(\vx, \lambda) \label{eqn:dual}. 
\end{align}

The {\it dual problem} is then given in below.

\begin{problem}[Dual problem]
	\label{prob:problem_4}
	\begin{equation}
		L^{\star}:=\min_{\lambda\geq 0} L(\lambda)
		\label{eqn:problem_4} 
	\end{equation}
\end{problem}

Applying the {\it Lagrangian optimality}, 
the feasible fraction of allocated frequencies $\vx^{\ast}$ that maximises the Lagrangian $L(\vx, \lambda)$ over $\vx\in\set{X}$ is given by
\begin{equation}
	x_s^{\ast}(\lambda) =
	\begin{cases}
		D_s,  \mbox{ if } \lambda\leq -\beta_s C_s &\\
 		\left(\min\left\{\dfrac{\alpha_s}{\lambda  +\beta_s C_s}-\dfrac{1}{A_s}, D_s\right\}\right)^{+},  \mbox{o.w.} &
	\end{cases}
	\label{eqn:solveX}
\end{equation}

The solution in \eqref{eqn:solveX} is derived by utilising that for any $\lambda$, 
the partial derivative $\partial L(\vx,\lambda)/\partial x_s = \alpha_s A_s/(1 + A_s x_s) -\beta_s C_s -\lambda$ is monotone decreasing over $x_s\in [0, D_s]$. 
If $\partial L(\vx,\lambda)/\partial x_s|_{x_s = D_s}\geq 0$, 
the partial derivative is non-negative at each point $x_s \in [0, D_s]$. 
Thus, $x_s = D_s$ maximises $L(\vx,\lambda)$ over $x_s\in[0, D_s]$ for any $\lambda\leq \alpha_s A_s/(1+A_sD_s) -\beta_s C_s$.
 Along similar lines, 
 we have that $x_s^{\ast}(\lambda) = 0$, 
 if $\partial L(\vx,\lambda)/\partial x_s|_{x_s = 0}\leq 0$,
 while $x_s^{\ast}(\lambda) = \alpha_s/(\lambda  +\beta_s C_s)-1/A_s$ 
 if $\partial L(\vx,\lambda)/\partial x_s|_{x_s = 0} \geq 0$ and $\partial L(\vx,\lambda)/\partial x_s|_{x_s = D_s} \leq 0$. 
 Moreover, for $ -\beta_s C_s <\lambda \leq  \alpha_s A_s/(1 + A_s x_s) -\beta_s C_s$, 
 we have that $\alpha_s/(\lambda  +\beta_s C_s)-1/A_s\geq D_s$, 
 and thus the second case in \eqref{eqn:solveX} also returns $x_s^{\ast}(\lambda) = D_s$ for $-\beta_s C_s < \lambda \leq  \alpha_s A_s/(1 + A_s x_s) -\beta_s C_s$. 
 As a result, the solution can be written in the form of~\eqref{eqn:solveX}.
 
The {\it complementary slackness} provides $\lambda^{\ast}(X_{\max}-\sum_s x_s^{\ast}) = 0$. 
Thus, $\sum_s x_s^{\ast} = X_{\max}$ if $\lambda^{\ast}>0$.

The solution to~\eqref{eqn:problem_4} can be given by numerically minimising $L(\lambda)$ over $\lambda\geq 0$. 
For this, we use a subgradient-based search, 
and update $\lambda$ iteratively by
\begin{equation}
	\lambda(t+1) = \left(\lambda(t) -\kappa(t) \left(X_{\max}-\sum_s x_s^{\ast}(\lambda(t))\right)\right)^{+}
	\label{eqn:updateLambda}
\end{equation}
where $x_s^{\ast}(t)$ is updated by \eqref{eqn:solveX} and $\kappa(t)$ is the step size at iteration $t$. 

The algorithm of iteratively performing \eqref{eqn:updateLambda} and \eqref{eqn:solveX} converges,
if $\kappa(t)$ is chosen appropriately~\cite[Section 6.3.1]{bertsekas1999nonlinear}. 
There are many ways to select the step size. 
For constant step size, 
the subgradient algorithm is guaranteed to converge within some range of the optimal value~\cite{boyd2003subgradient}.
Note that, from~\eqref{eqn:solveX}, 
if $\lambda\geq \max_s (\alpha_s A_s-\beta_s C_s)$, 
no service will be scheduled. 
Therefore, the optimal $\lambda$ lies in the interval $[0, \max_s (\alpha_s A_s-\beta_s C_s)]$. 
In order to provide a good starting point for the fast convergence to the solution, 
we can numerically minimise the univariate function $L(\lambda)$ over a finite set of uniformly distributed $\lambda$ in $[0, \max_s (\alpha_s A_s-\beta_s C_s)]$,  
and choose the $\lambda$ corresponding to the minimum value of $L(\lambda)$ as the starting point.

Given the optimal $\lambda^{\ast}$ and the corresponding $x_s^{\ast}(\lambda^{\ast})$ for all $s\in\Ss^{(d')}$, 
the solution to the equivalent problem~\eqref{eqn:OptUtility_prob_2} can be obtained by collecting $x_s^{\ast}(\lambda^{\ast})$ for $s\in\Ss^{(d')}$ and $p_s = 0$ for $s\in\Ss\setminus\Ss^{(d')}$ in the joint vector $\vp'$. 

\begin{figure}[t]
  \centering
	\includegraphics[width=.43\textwidth]{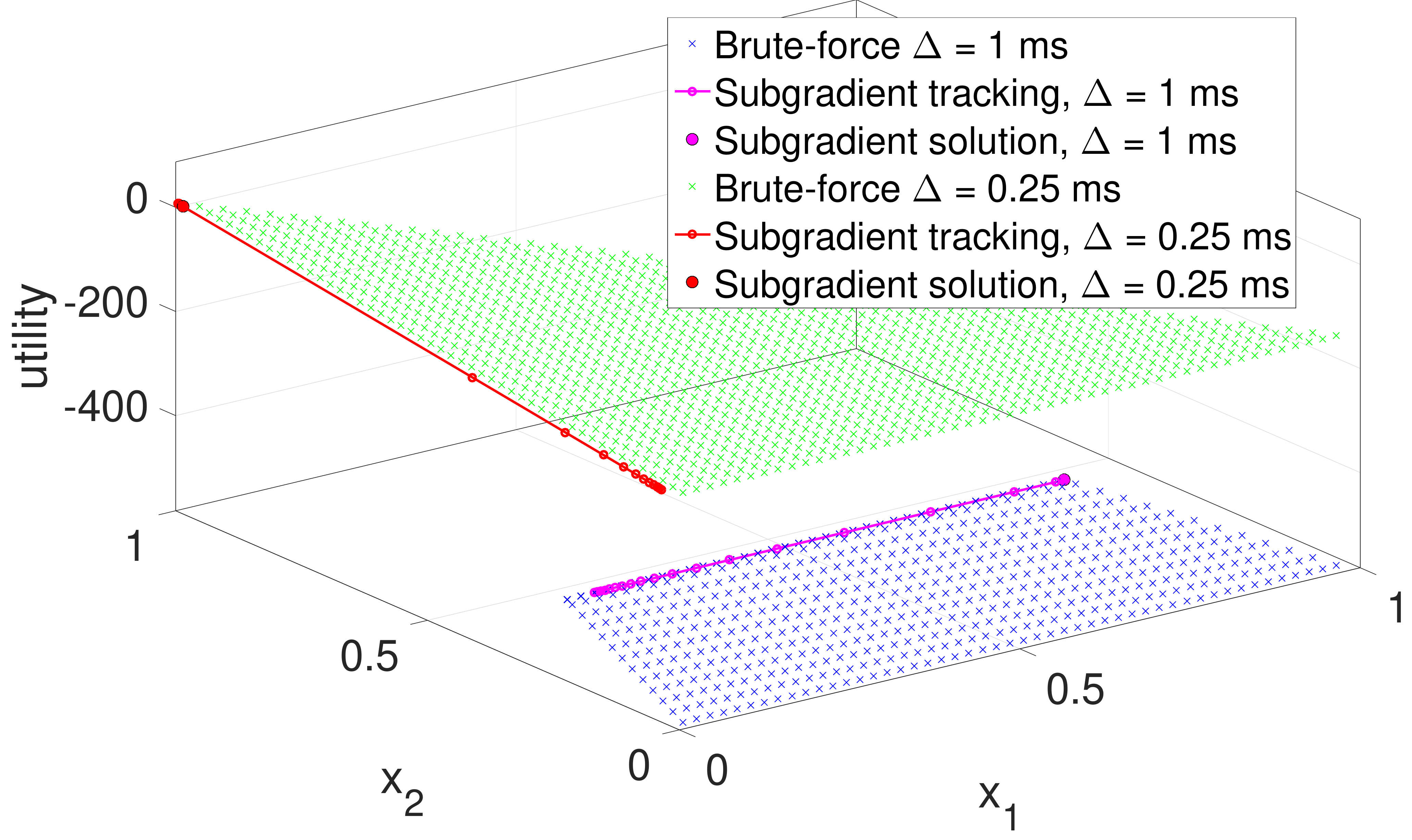}
  \caption{Subgradient-based search over $\{\vx: x_s\in[0,D_s], \forall s; \sum_s x_s\leq X_{\max}\}$ for Example \ref{exam:costFunct}. $\alpha_1 = 0.9, \alpha_2 = 0.85$, $\beta_1 = 0.1, \beta_2 = 0.15$, $\kappa = 1$.}
  \label{fig:SubgradientSearch}
\end{figure}

Fig.~\ref{fig:SubgradientSearch} shows that the subgradient-based searching algorithm based on \eqref{eqn:solveX} and \eqref{eqn:updateLambda} converges to the optimal solution $x_1 = 0, x_2 = 1$ for the previous introduced Example~\ref{exam:costFunct}. 
Small $\beta$s are chosen because the cost function has a much larger scale than the concave utility of rate in this example. 
We have $\beta_2>\beta_1$ because user $2$ has higher latency requirement.
Using $\Delta = 0.25$\,ms achieves generally higher utility than using $\Delta = 1$\,ms 
because the latter violates the latency constraint and results in a higher cost. 
The best solution with respect to $\Delta = 1$\,ms is $x_1 = 0.75, x_2 = 0.25$.  

\section{Numerical Results}
\label{num_results}

In this section, we analyse the performance of the proposed user-centric scheduler, 
by jointly considering the delay of the MCC services and the throughput of the MBB services. 
In detail, we compare our proposed scheme employing the set $\Ts=\{0.1~{\rm\,ms}, 0.2~{\rm\,ms}, \ldots, 1~{\rm\,ms}\}$ of scalable TTI lengths against a baseline scheduler optimising the duplexing mode and the resource allocation between users to maximise the same utility \eqref{eqn:utilityDef} but with fixed TTI lengths. 
The tradeoff between the performance of MCC and MBB services can be flexibly tuned with parameters ${\alpha^{\rm (MBB)}, \beta^{\rm (MBB)}, \alpha^{\rm (MCC)}, \beta^{\rm (MCC)}}$, which denote the service-specific weight factors $\alpha_s$ and $\beta_s$ for MBB and MCC services, respectively. 
The joint selection of ${[\alpha_s, \beta_s]}$ allows the scheduler to perform a truly user-centric allocation of the available resources. 
In the following results, reasonable values of $\alpha_s$ and $\beta_s$ have been selected for the two main service categories, MBB and MCC.

 
We focus our analysis on the isolated pico BS scenario defined in~\cite[Sect. 6.2]{3gpp_tr36828} 
where the transmit power of both BSs and UEs are set to have an average signal to noise ratio at the cell edge of 10~dB. 
All the other simulation parameters mainly related to  line-of-sight probability, path-loss and shadowing can be found in~\cite[Tab. 6.2-1]{3gpp_tr36828}. 
Moreover, we assume that the BS serves 
2 MBB UEs active in the DL, modelled as full buffer UEs, 
and 10 MCC UEs active in UL, which generate small packets of size 125 bytes~\cite{metis_d11}, 
whose arrival rates are Poisson distributed with parameter $\eta$.
Then, we further assume that the control signal transmission requires $\tau=0.05$\,ms.

\begin{figure}[t!]
\centering
                \begin{subfigure}[b]{0.4\textwidth}
                \includegraphics[width=\textwidth]{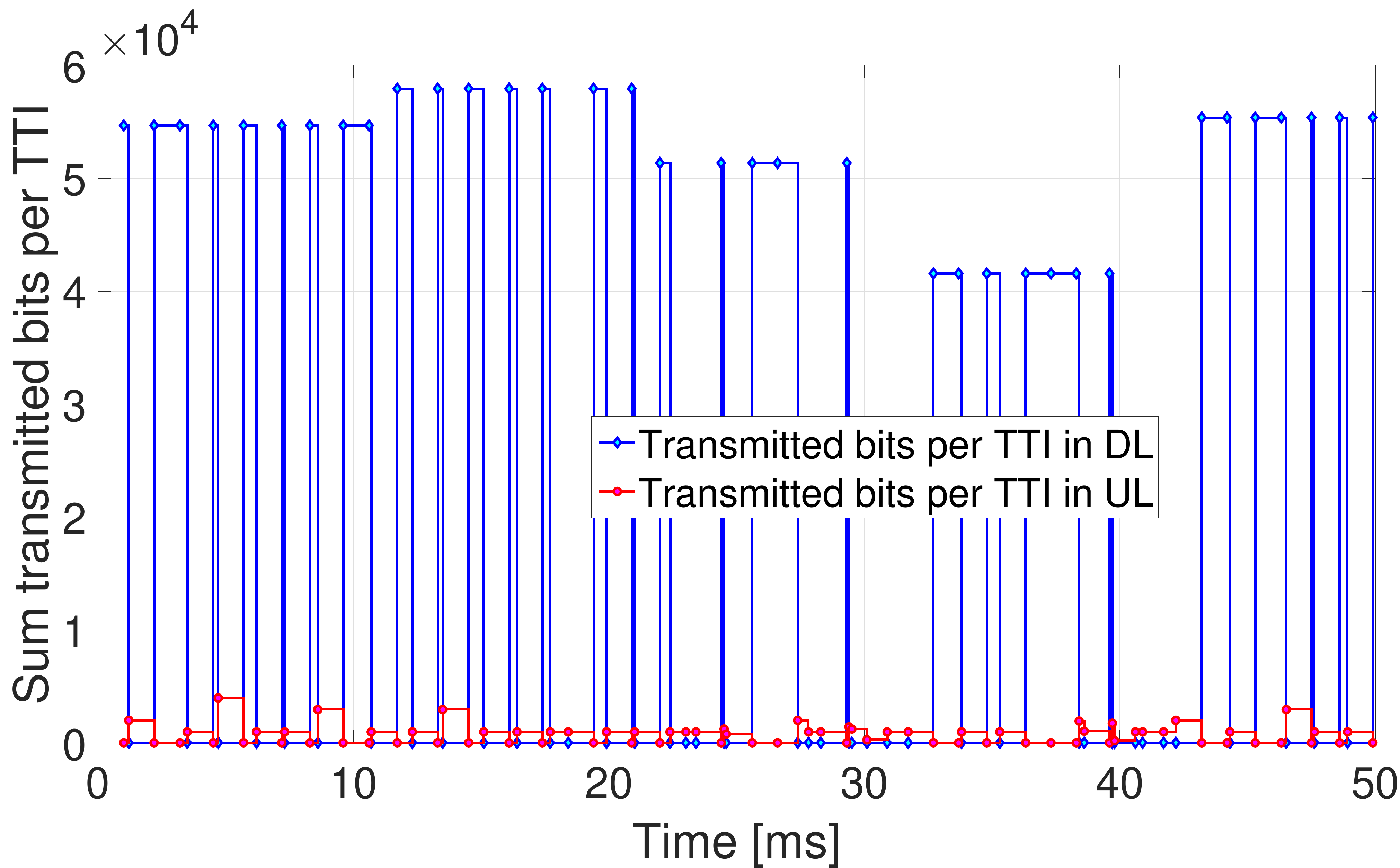}
                \caption{Adaptation of duplexing with $\eta = 100$ packet/s..}
                \label{fig:TransBit_lambda100}
        \end{subfigure}
                \begin{subfigure}[b]{0.4\textwidth}
                \includegraphics[width=\textwidth]{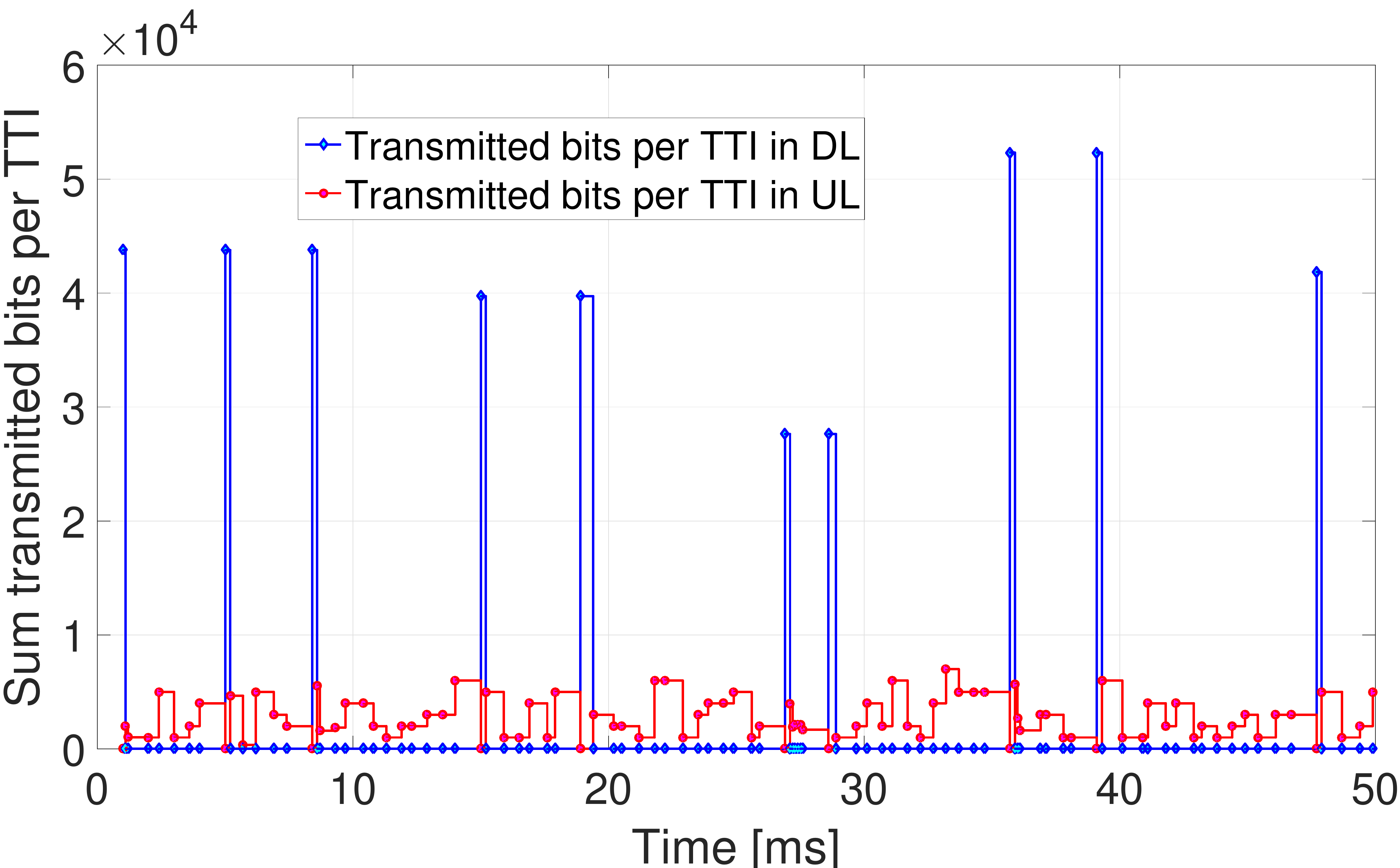}
                \caption{Adaptation of duplexing with $\eta = 500$ packet/s.}
                \label{fig:TransBit_lambda500}
        \end{subfigure}
        %
        \caption{Dynamic duplexing and TTI length adapted to packet arrival with $\alpha^{\rm (MBB)} = 1$, $\beta^{\rm (MBB)} = 0.2$, $\alpha^{\rm (MCC)} = 0.01$, $\beta^{\rm (MCC)}=3$, and $c_{\min} = 10^{-6}$.}
        \label{fig:Adaptation_Duplexing}
\end{figure}

{\it 1) Adaptation of duplexing, TTI length and scheduled services over time.} 
In Fig.~\ref{fig:Adaptation_Duplexing}, 
we report the number of bits transmitted over time 
to show how our proposed algorithm is able to adapt to the MCC packet arrival, 
well selecting both the duplexing (UL or DL) and the TTI length. 

By setting a high value for $\beta^{\rm (MCC)}$, e.g., $\beta^{\rm (MCC)}=3$, we increase the priority of MCC services 
and more resources are allocated to them in UL. As a consequence, when we increase the packet arrival rate from $\eta=100$ packet/s in Fig.~\ref{fig:TransBit_lambda100} to $\eta=500$ packet/s in Fig.~\ref{fig:TransBit_lambda500}, we observe that more UL TTIs are scheduled in order to fulfil the latency requirements of the MCC services, causing less resource allocation to MBB services in DL.



%

%

\begin{figure}[t]
\centering
 	\includegraphics[width=0.40\textwidth]{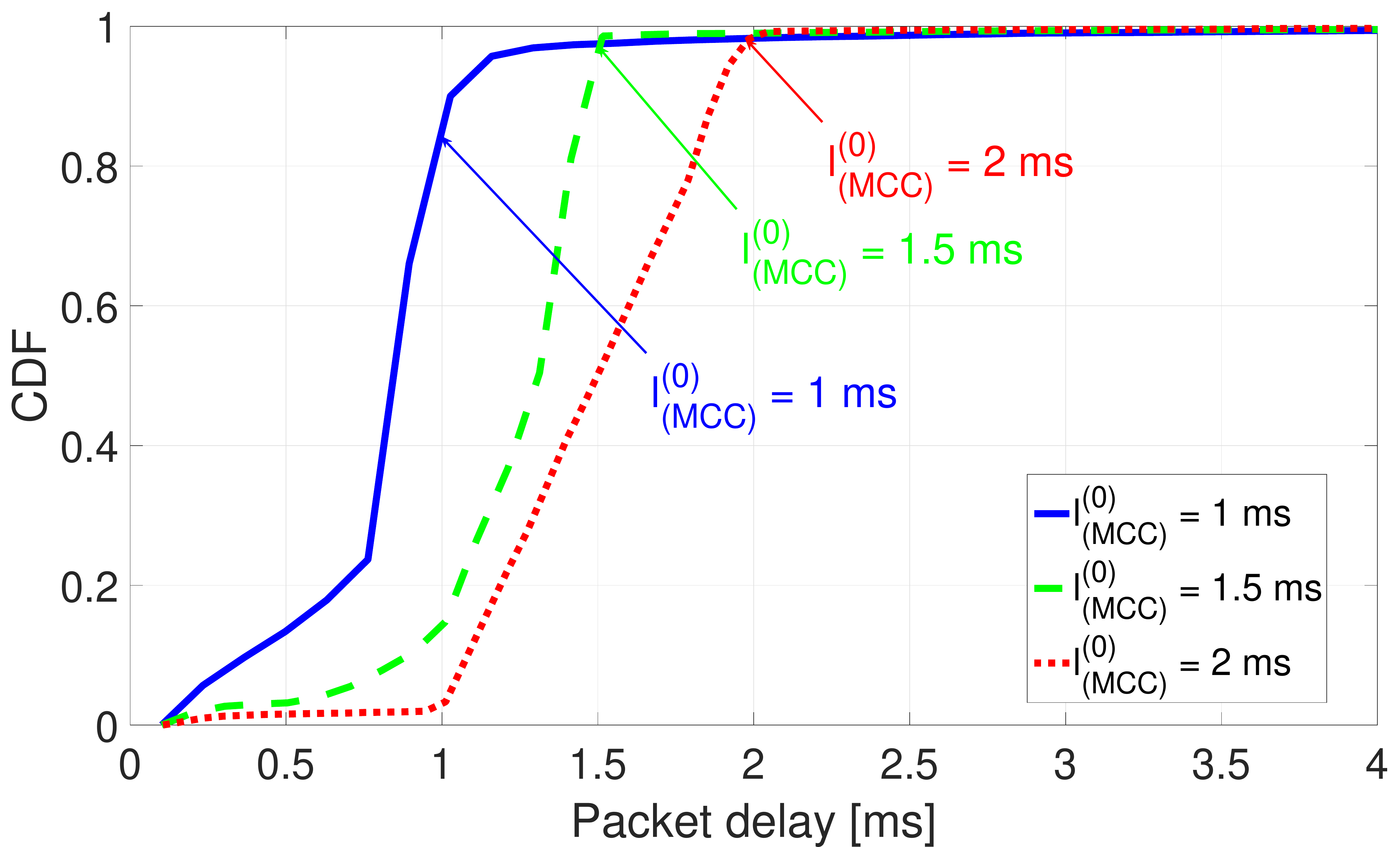}
       \caption{CDF of delay of MCC packets depending on predefined latency constraints with 
       $\eta=100$\,packet/s, $\alpha^{\rm (MBB)} = 1$, $\beta^{\rm (MBB)} = 0.2$, $\alpha^{\rm (MCC)} = 0.01$, $\beta^{\rm (MCC)} = 3$, and $c_{\min} = 10^{-6}$.}
        \label{fig:latencyConstraint}
\end{figure}

{\it 2) Predefining latency requirements for MCC services.}
For each MCC user/service-specific requirement, 
we can define a specific latency constraint (see also~\eqref{eqn:remainLatency}).  
Fig.~\ref{fig:latencyConstraint} shows the cumulative distribution function (CDF) of the delay of the MCC packets for different values of the latency constraint: 
these results show that the proposed resource allocation scheme manages to meet this latency constraint in more than 98\,\% of the cases when it is larger than 1\,ms. For a more strict latency constraint, e.g. 1\,ms, the scheme meets it in almost 90\,\% of the cases: this happens because a service cannot be scheduled before the beginning of the next TTI. For instance, in case a service arrives within the first 0.1\,ms of the $n$th TTI and $\Delta(n) = 1$\,ms, it has to wait for more than 0.9\,ms to be scheduled. Even if the shortest length $\Delta(n+1) = 0.1$\,ms is chosen for the next TTI, the latency constraint of 1\,ms cannot be met. 


%
%
%

\begin{figure}[t]
\centering
 \includegraphics[width=0.42\textwidth]{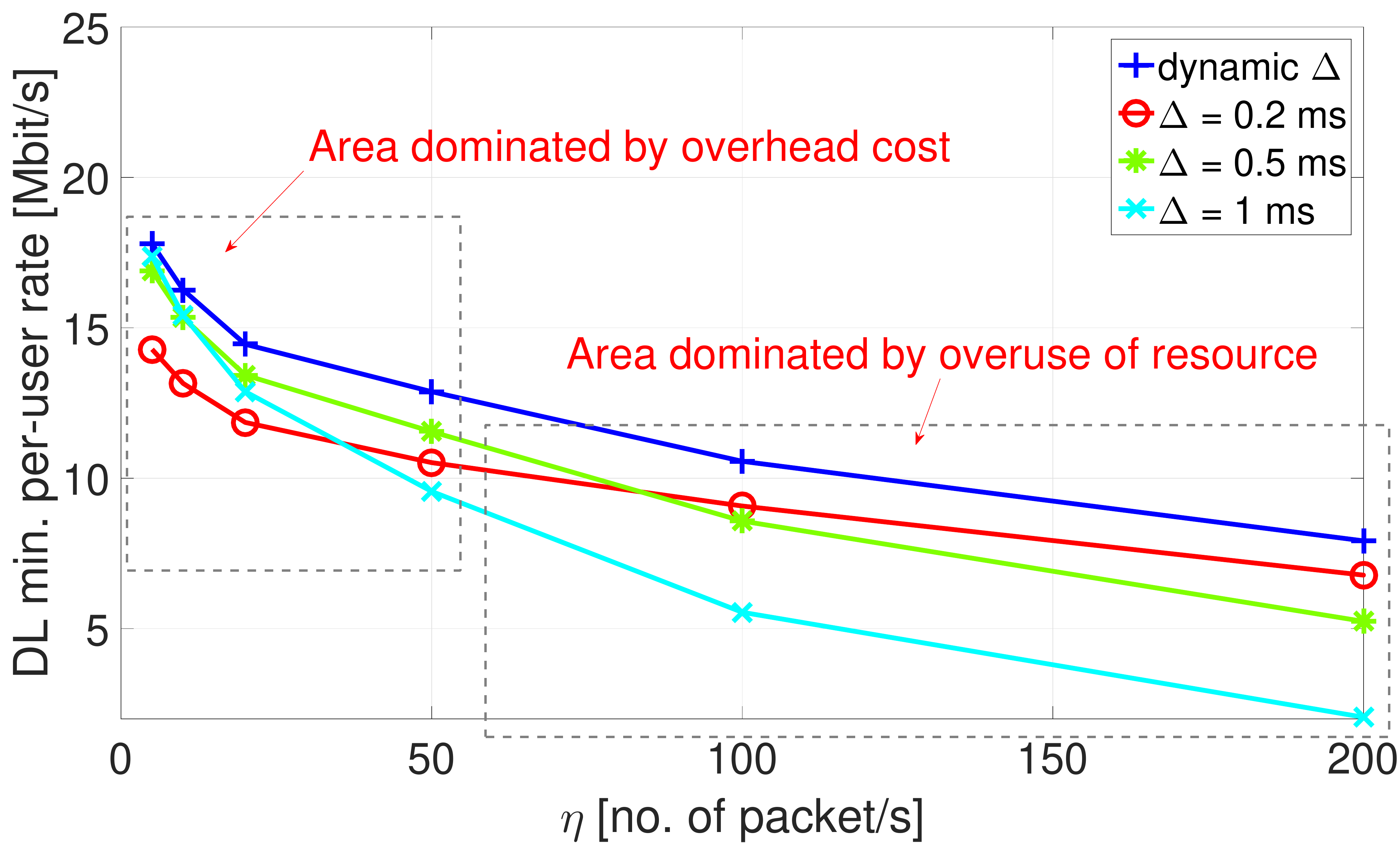}
                \caption{Minimum average MBB user throughput for different values of $\eta$, $\beta^{\rm (MCC)}=3$.}
        \label{fig:RateRegion}
\end{figure}

{\it 3) Inappropriate selection of fixed TTI length leads to capacity loss.} 
Assuming that high priority is given to the MCC services, 
the throughput of MBB services in DL mainly depends on the following two factors: 
\emph{a)} the overhead cost $\psi(\Delta)$, and 
\emph{b)} the probability that there exist MCC services within the duration of a TTI.
This probability is denoted by $P(\Delta) := \Pr\{N(t)>0|t \in[t', t' + \Delta]\}$, 
where $N(t)$ is the number of MCC services at time $t$, 
and $t'$ is an arbitrary time instant. 
It is important to note that a longer TTI length $\Delta$ reduces the throughput loss due to the proportionally less overhead related to the control signal transmission.
However, as the packet arrival rate  $\eta$ increases, 
larger $\Delta$ also causes much higher $P(\Delta)$, 
thus longer time periods will be occupied by the MCC services than actually needed. 
For instance, 
if 10\,MCC packets arrive uniformly on a time interval of $10$\,ms, 
then with $\Delta = 1$\,ms, 
all slots may be allocated to MCC services and leave the MBB services with no resources, 
while by configuring $\Delta = 0.2$\,ms, 
only $2$\,ms will be allocated to the 10\,MCC packets, 
and the remaining $8$\,ms can be allocated to MBB services. 
In Fig.~\ref{fig:RateRegion}, 
we report the minimum MBB user rate for different values of $\eta$ and show that fixed TTI length cannot cope with the tradeoff between $\psi(\Delta)$ and $P(\Delta)$, 
while our proposed scheme with dynamic TTI adaptively finds a good tradeoff and provides an enhanced throughput performance to MBB services.

\begin{figure}[t]
\centering
                \begin{subfigure}[b]{0.42\textwidth}
                \includegraphics[width=\textwidth]{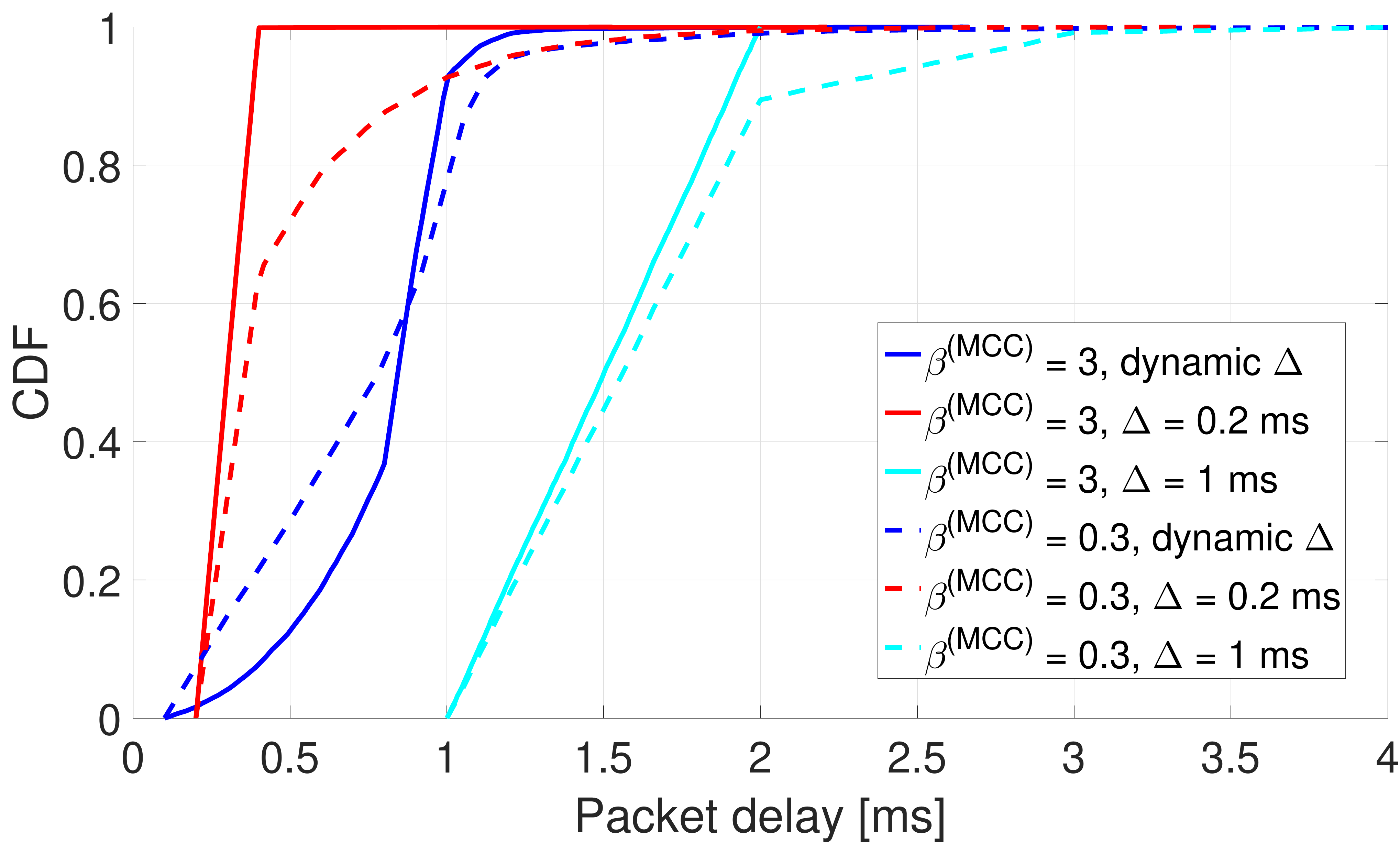}
                \caption{CDF of delay of MCC packets, $\eta = 300$ packet/s.}
                \label{fig:Delay_Comparison_BetaMCC}
        \end{subfigure}
               \begin{subfigure}[b]{0.42\textwidth}
                \includegraphics[width=\textwidth]{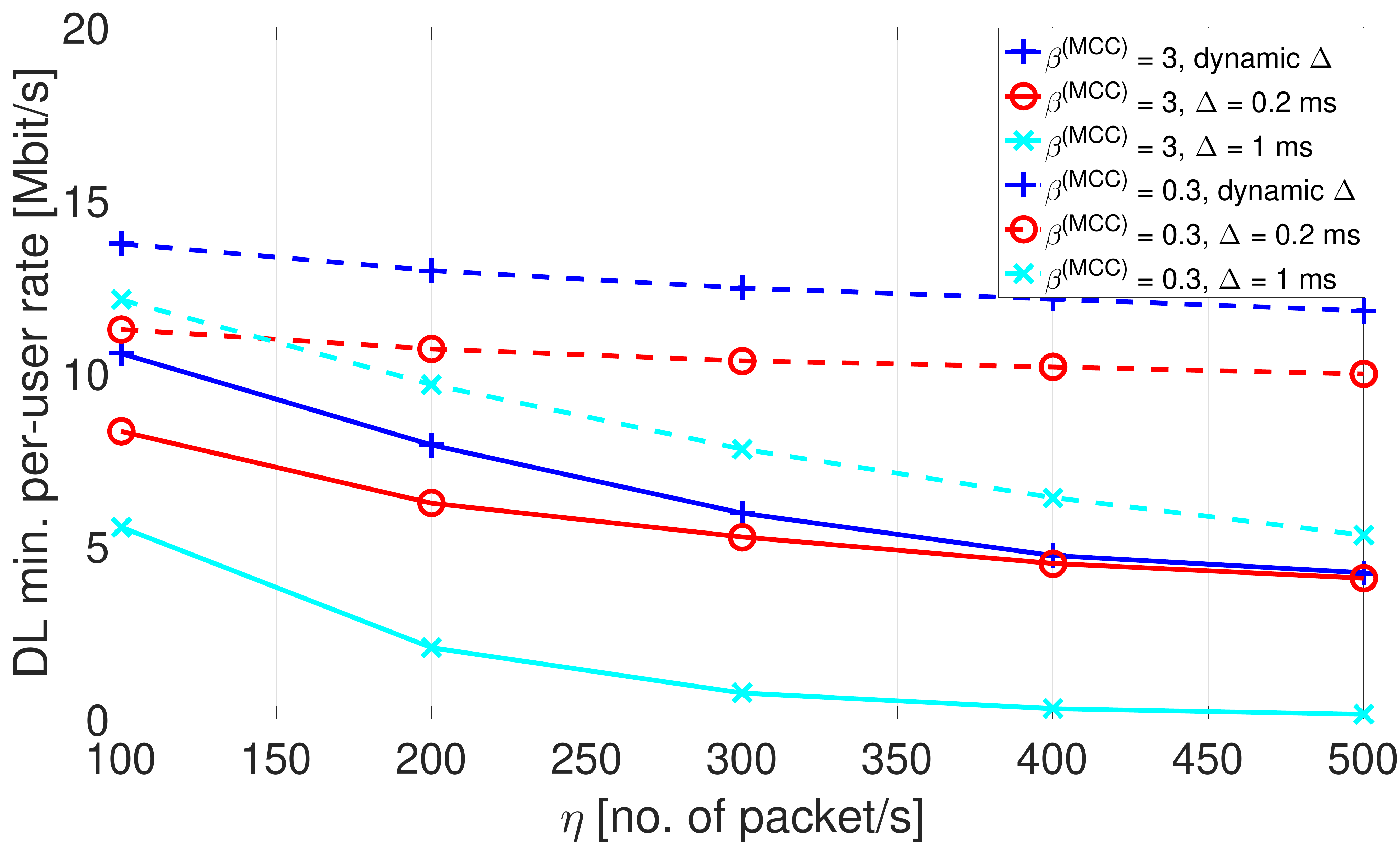}
                \caption{Minimum average MBB user throughput for different values of $\eta$.}
                \label{fig:Rate_Comparison_BetaMCC}
        \end{subfigure}
\caption{Tradeoff between delay of MCC packets and throughput of MBB users obtained by tuning $\beta^{\rm (MCC)}$, $l_{(MCC)}^{(0)} = 1$\,ms.}
\label{fig:BetaMCC_Impact}
\end{figure}

{\it 4) Flexible tradeoff between delay of MCC services and throughput of MBB services by tuning parameters.} 
From Fig.~\ref{fig:RateRegion}, 
we notice that by giving high priority to MCC services to fulfil their latency requirements, 
we sacrifice the throughput of MBB services as $\eta$ increases. 
However, if we aim to provide an enhanced throughput performance to MBB services, 
while accepting a slight performance degradation on delay of MCC services, 
we can reduce $\beta^{\rm (MCC)}$ (or, similarly, increase $\alpha^{\rm (MBB)}$). 
Fig.~\ref{fig:BetaMCC_Impact} shows the CDF of delay of the MCC packets and the minimum MBB user rate for different values of $\beta^{\rm (MCC)}$.
First of all, we observe that although the baseline scheme with fixed $\Delta = 0.2$\,ms is able to guarantee a lower delay to the MCC services 
when compared to our scalable TTI proposal (see Fig.~\ref{fig:Delay_Comparison_BetaMCC}), 
it also strongly loses in the MBB user rate (see Fig.~\ref{fig:Rate_Comparison_BetaMCC}), 
mainly because of the higher impact of the control signalling overhead with very short TTI length.
Moreover, by selecting a low $\beta^{\rm (MCC)} = 0.3$, 
the system achieves an operating point with robust throughput of MBB services 
and acceptable slightly longer delays of MCC services with respect to the case with $\beta^{\rm (MCC)} = 3$. It is also worth mentioning that the proposed scheme with scalable TTI significantly outperforms the fixed $\Delta = 1$\,ms  (configured TTI length in LTE) in terms of both delay and throughput. 

Fig.~\ref{fig:BetaMCC_Impact} allows us also to quantify better the benefits of the proposed scheme with scalable TTI against the baseline with fixed TTI length. 
For example, by comparing our proposal against a scheme with fixed TTI of 0.2\,ms and 1\,ms respectively for $\eta=300$ packet/s and $\beta^{\rm (MCC)} = 0.3$, 
we observe that both the scalable TTI and fixed $\Delta = 0.2$\,ms provide comparable performance of latency below 2\,ms to all the MCC users, while the fixed $\Delta = 1$\,ms provides the same latency to only $90$\% of the MCC users.
However, the scalable scheme provides $20$\% gain in the average MBB user throughput when compared to fixed $\Delta = 0.2$\,ms and $40$\% gain when compared to fixed $\Delta = 1$\,ms.
Similarly, for a target MBB user throughput of about 12\,Mbps, 
the baseline scheme with fixed $\Delta = 0.2$\,ms can support up to only $\eta=100$\,packet/s for the MCC users, 
whereas our dynamic proposal can cope with more than $\eta\geq 300$\,packet/s, 
providing a significant gain in the number of MCC packets that can be served with latency below 2\,ms while simultaneously supporting MBB user throughput of 12\,Mbps.

\section{Conclusions} \label{conclusions}

In this paper, 
in order to cope with mixed traffic types,
we have presented a new user-centric scheduling approach 
based on a  dynamic TDD framework and with flexible TTI length configuration capabilities. 
In this framework, 
we have defined service-specific weight factors ($\alpha_s$, $\beta_s$) to better address the heterogeneous rate or latency requirements characterising each user. 
The optimisation variables of our scheduling scheme are the selection of UL or DL direction, the TTI length and the services to be scheduled.
Extensive simulations show the remarkable performance gains of the proposed scheduling approach with respect to one with fixed TTI lengths  
in terms of both rate achieved by the MBB users and latency guaranteed to the MCC users.


\bibliographystyle{IEEEtran}
\bibliography{refs}

\begin{thebibliography}{10}
\providecommand{\url}[1]{#1}
\csname url@samestyle\endcsname
\providecommand{\newblock}{\relax}
\providecommand{\bibinfo}[2]{#2}
\providecommand{\BIBentrySTDinterwordspacing}{\spaceskip=0pt\relax}
\providecommand{\BIBentryALTinterwordstretchfactor}{4}
\providecommand{\BIBentryALTinterwordspacing}{\spaceskip=\fontdimen2\font plus
\BIBentryALTinterwordstretchfactor\fontdimen3\font minus
  \fontdimen4\font\relax}
\providecommand{\BIBforeignlanguage}[2]{{%
\expandafter\ifx\csname l@#1\endcsname\relax
\typeout{** WARNING: IEEEtran.bst: No hyphenation pattern has been}%
\typeout{** loaded for the language `#1'. Using the pattern for}%
\typeout{** the default language instead.}%
\else
\language=\csname l@#1\endcsname
\fi
#2}}
\providecommand{\BIBdecl}{\relax}
\BIBdecl

\bibitem{ngmn_whitepaper}
NGMN, ``{NGMN 5G white paper},'' {Next generation mobile networks (NGMN)}, A
  deliverable by the NGMN Alliance, Feb. 2015.

\bibitem{pedersen_cm16}
K.~I. Pedersen, G.~Berardinelli, F.~Frederiksen, P.~Mogensen, and A.~Szufarska,
  ``A flexible {5G} frame structure design for frequency-division duplex
  cases,'' \emph{IEEE Communications Magazine}, vol.~54, no.~3, pp. 53--59,
  Mar. 2016.

\bibitem{durisi_arxiv16}
G.~Durisi, T.~Koch, and P.~Popovski, ``Towards massive, ultra-reliable, and
  low-latency wireless communication with short packets,''
  http://arxiv.org/abs/1504.06526, Mar. 2016.

\bibitem{lahetkangas_icc14}
E.~L{\"{a}}hetkangas, K.~Pajukoski, J.~Vihri{\"{a}}l{\"{a}}, G.~Berardinelli,
  M.~Lauridsen, E.~Tiirola, and P.~Mogensen, ``Achieving low latency and energy
  consumption by {5G TDD} mode optimization,'' in \emph{Proc. IEEE
  International Conference on Communications (ICC)}, Sydney (Australia), Jun.
  2014.

\bibitem{3gpp_tr36828}
3GPP, ``{Further enhancements to LTE Time Division Duplex (TDD) for
  Downlink-Uplink (DL-UL) interference management and traffic adaptation
  (Release 11)},'' {3rd Generation Partnership Project (3GPP)}, TR {36.828},
  Jun. 2012.

\bibitem{ding_icc14}
M.~Ding, D.~Lopez-Perez, A.~Vasilakos, and W.~Chen, ``Dynamic {TDD}
  transmissions in homogeneous small cell networks,'' in \emph{Proc. IEEE
  International Conference on Communications (ICC)}, Sydney (Australia), Jun.
  2014.

\bibitem{baracca_vtc16}
P.~Baracca, ``Traffic profile based clustering for dynamic {TDD} in dense
  mobile networks,'' in \emph{Proc. IEEE Vehicular Technology Conference (VTC
  Fall)}, Montreal (Canada), Sep. 2016.

\bibitem{pedersen2015flexible}
K.~Pedersen, F.~Frederiksen, G.~Berardinelli, and P.~Mogensen, ``A flexible
  frame structure for {5G} wide area,'' in \emph{Proc. Vehicular Technology
  Conference (VTC Fall)}, Boston (MA), Sep. 2015.

\bibitem{levanen2014radio}
T.~Levanen, J.~Pirskanen, and M.~Valkama, ``Radio interface design for
  ultra-low latency millimeter-wave communications in {5G} era,'' in
  \emph{Proc. IEEE Globecom Workshops (GC Wkshps)}, Austin (TX), Dec. 2014.

\bibitem{boyd2004convex}
S.~Boyd and L.~Vandenberghe, \emph{Convex optimization}.\hskip 1em plus 0.5em
  minus 0.4em\relax Cambridge university press, 2004.

\bibitem{bertsekas1999nonlinear}
D.~P. Bertsekas, \emph{Nonlinear programming}.\hskip 1em plus 0.5em minus
  0.4em\relax Athena scientific, 1999.

\bibitem{boyd2003subgradient}
S.~Boyd, L.~Xiao, and A.~Mutapcic, ``Subgradient methods,'' \emph{lecture notes
  of EE392o, Stanford University, Autumn Quarter}, 2003.

\bibitem{metis_d11}
``Scenarios, requirements and {KPIs} for {5G} mobile and wireless system,''
  METIS D1.1, Apr. 2013.

\end{thebibliography}
		
\end{document}